\documentclass{article}
\usepackage{spconf,amsmath,graphicx,amssymb,amsfonts}
\usepackage[utf8]{inputenc}
\usepackage{setspace}
\usepackage{bbm}
\usepackage{algorithmic}
\usepackage{algorithm}
\usepackage{cite,color}
\usepackage{bm,bbold}
\usepackage{url}
\usepackage[title]{appendix}
\newcommand\bw{\ensuremath{{\bm w}}}
\newcommand\bW{\ensuremath{{\bm W}}}

\newcommand\bx{\ensuremath{{\bm x}}}
\newcommand\by{\ensuremath{{\bm y}}}
\newcommand\bG{\ensuremath{{\bm G}}}
\newcommand\bh{\ensuremath{{\bm h}}}

\newcommand\bz{\ensuremath{{\bm z}}}

\newcommand\cO{\ensuremath{\mathcal{O}}}

\newcommand\bA{\ensuremath{{\bm A}}}

\newcommand\bb{\ensuremath{{\bm b}}}

\newcommand\bu{\ensuremath{{\bm u}}}

\newcommand{\Rbb}{\mathbb{R}}
\newcommand{\Cbb}{\mathbb{C}}

\newcommand{\setX}{\mathcal{X}}

\newcommand{\setC}{\mathcal{C}}

\newcommand{\jj}{\mathfrak{j}}

\newcommand\bQ{\ensuremath{{\bm Q}}}

\newcommand{\bI}{{\bm I}}

\newtheorem{Fact}{Fact}

\title{AN EFFICIENT ALGORITHM FOR MULTIUSER SUM-RATE MAXIMIZATION OF LARGE-SCALE Active RIS-AIDED
MIMO SYSTEM}

\name{Qian Zhang$^{\dag}$, Mingjie Shao$^{\dag}$, Qiang Li$^{\S}$  and Ju Liu$^{\dag}$
\thanks{\scriptsize This work was supported in part by the National Natural Science Foundation of China under Grant 62071275 and Grant 62171110; in part by the Key R\&D Plan of Shandong Province of China under Grant2021SFGC0701. 
 Corresponding authors: Mingjie Shao (mingjieshao@sdu.edu.cn) and Ju Liu (juliu@sdu.edu.cn).}}
\address{\small $^{\dag}$School of Information Science and Engineering, Shandong University, Qingdao, China\\
\small $^{\S}$School of Information and Communication Engineering, University of Electronic Science and Technology of China, Chengdu, China}

\begin{document}

\maketitle
\ninept

\begin{abstract}
Active reconfigurable intelligent surface (RIS) is a new RIS architecture that can reflect and amplify communication signals. 
It can provide enhanced performance gain compared to the conventional passive RIS systems that can only reflect the signals.
On the other hand, the design problem of active RIS-aided systems is more challenging than the passive RIS-aided systems and its efficient algorithms are less studied.
In this paper, we consider the sum rate maximization problem in the multiuser massive multiple-input single-output (MISO) downlink with the aid of a large-scale active RIS.
Existing approaches for handling this problem usually resort to general optimization solvers and can be computationally prohibitive. 
We propose an efficient block successive upper bound minimization (BSUM) method, of which each step has a (semi) closed-form update. 
Thus, the proposed algorithm has an attractive low per-iteration complexity.
By simulation, our proposed algorithm consumes much less computation than the existing approaches. 
In particular, when the MIMO and/or RIS sizes are large, our proposed algorithm can be orders-of-magnitude faster than existing approaches. 

 \end{abstract}

\section{Introduction}

Reconfigurable intelligent surface (RIS) refers to a planar of adaptive electromagnetic elements that can be designed to improve the wireless channel environment \cite{zhang2018metasurfaces,ren2020metasurface,venkatesh2020metasurface}.
When it was first proposed, the RIS elements could passively reflect the signal such that system performance metrics such as multiuser sum rate \cite{guo2020rissumrate}, energy efficiency \cite{huang2019risenergy}, and symbol-error probability \cite{shao2020symbolerror,liu2021symbollevel} are enhanced.
However, it is also found that the reflected signal power of passive RIS-aided systems is limited when the RIS size is not large enough or when the transmission distance is long \cite{zhang2023active}.
To alleviate this limitation, recently the active RIS architecture was proposed \cite{long2021active,zhang2023active,you2021active}.
Compared to the passive RIS systems, the active RIS systems are equipped with active amplifying components in order to reflect and amplify the signal at the same time \cite{long2021active,zhang2023active}.
It has been reported that under the same total power budget of the base station (BS) and RIS, active RIS-aided systems can provide superior system gain than passive RIS-aided systems \cite{zhang2023active}.
We refer the readers to \cite{zhang2023active,you2021active} for a more detailed discussion of the active and passive RIS systems.

On the other hand, the design problem of active RIS systems poses a different form from that of passive RIS systems, which calls for new efficient design algorithms.
The most widely adopted approach in the literature is to apply general purpose optimization solvers, such as CVX \cite{grant2014cvx}, to solve the design problems in a RIS-aided system; see, e.g., sum rate maximization \cite{zhu2022subconnect}, mean square error minimization \cite{allu2023cognitive}, and energy efficiency maximization \cite{liu2022subconnect}.
In \cite{zhu2022subconnect}, Zhu {\it et al.} tackled the sum rate maximization problem by a dedicated algorithm that composes block coordinate descent (BCD), alternating direction method of multipliers (ADMM), CVX, and majorization-minimization (MM), which is a three-tier iterative algorithm. 
In future 6G systems, the BS will be associated with massive MIMO systems, which may contain hundreds of antennas; the number of RIS elements can also be large in order to support high performance. 
These physical aspects will contribute to a large-scale joint design problem for beamforming in active RIS-aided systems. 
The aforementioned methods can inevitably incur unaffordable computational complexity because the computational complexity of general purpose solvers scales quickly with the problem dimensions. 

In this paper, we propose an efficient algorithm for active RIS-aided multiuser multiple-input single-output (MISO) downlink systems. 
Our focus is multiuser sum rate maximization  (SRM) under the joint constraints arising from the BS and active RIS.
We apply the block successive upper bound minimization (BSUM) method \cite{razaviyayn2013unified} to the resulting problem. 
The BSUM algorithm breaks down the design problem by a recursive optimization process, and each subproblem involves only one variable block.
One key feature of our algorithm is that each step has a closed-form solution, either by an exact minimization or the proximal distance algorithm \cite{keys2019proximal}.
As a result, our proposed algorithm enjoys a low per-iteration complexity. 
Simulation results demonstrate that our algorithm is much faster than existing methods when attaining the same sum rate performance;
in particular, our algorithm can be over 1000 times faster when the MIMO and RIS sizes are large.

\vspace{-0.1cm}
\section{System Model}

\begin{figure}[t]	
	\centering \includegraphics[width=0.75\linewidth]{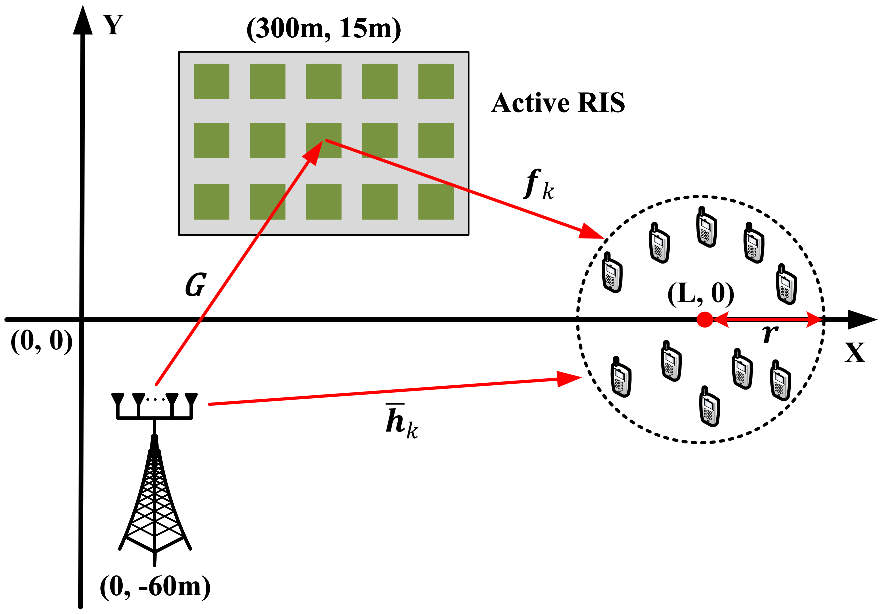}
	\caption{Illustration for single-cell active RIS-aided multiuser MIMO downlink scenario.}
	\label{fig:system_model}
\end{figure}

Consider a single-cell active RIS-aided multiuser multiple-input single-output (MISO) downlink scenario, which is shown in Fig.~\ref{fig:system_model}.
A base station (BS), which is equipped with $M$ antennas, employs linear precoding to serve $K$ single-antenna users with the aid of an active RIS equipping with $N$ reflecting elements.
Let $\boldsymbol{\overline{h}}_k \in \mathbb{C}^{M\times 1}$, $\boldsymbol{f}_k \in \mathbb{C}^{N\times 1}$, and $\bG \in \mathbb{C}^{N\times M}$ denote the downlink channels from BS to user $k$, RIS to user $k$, and BS to RIS, respectively.
Also, denote $\boldsymbol{w}_k \in \mathbb{C}^{M \times 1}$ as the linear precoding vector for the user $k$ at the BS and $\bm \phi\in \Cbb^{N \times 1}$ as the reflecting coefficients of active RIS elements.
Then, the achievable rate of user $k$ can be characterized as \cite{you2021active,zhang2023active}
\begin{equation}\label{eq:rate_k}
	\begin{split}
		R_k(\bw,\bm \phi ) = {\rm{log}}_2\, (1+{\sf SINR}_k), \enspace k = 1,\ldots, K,
	\end{split}
\end{equation}
where $\bw = [\boldsymbol{w}_1^T,\boldsymbol{w}_2^T,\dots,\boldsymbol{w}_K^T]^T$,
\[
{\sf SINR}_k = \frac{\vert \boldsymbol{h}_k^H \boldsymbol{w}_k \vert^2}{ \sum_{i=1,i\neq k}^{K} \vert \boldsymbol{h}_k^H \boldsymbol{w}_i \vert^2 + \Vert \boldsymbol{f}_k^H \bold{\Phi} \Vert_2^2 \sigma_v^2 + \sigma_k^2}
\]
is the signal-to-interference-plus-noise ratio (SINR) of user $k$; $\bm \Phi = \mbox{Diag}(\bm \phi)$ is a diagonal matrix with the diagonal entries being $\phi$; $\sigma_v^2$ and $\sigma_k^2$ denote the noise power at the active RIS and user sides, respectively; $\boldsymbol{h}_k^H = \boldsymbol{\overline{h}}_k^H + \boldsymbol{f}_k^H \bold{\Phi} \bG \in \mathbb{C}^{1\times M}$.

We jointly optimize the precoder $\bw$ and the reflecting coefficients $\bm \phi$ to maximize the multiuser sum-rate under both BS power constraints and active RIS power constraints.
The problem of interest can be formulated as
\begin{subequations}\label{eq:prob_form}
	\begin{align}
\min_{ \bw,\bm\phi }&~
		f(\bw,\bm\phi) = -\sum_{k=1}^{K} R_k(\bw,\bm\phi)  \\[-3pt]
		\mbox{s.t. }
&~\setC_{\sf BS}: ~ \sum_{k=1}^{K}\Vert  \boldsymbol{w}_k \Vert_2^2 \leq P_B,\label{eq:const1}\\[-3pt]
&~\setC_{\sf RIS}:~\vert \phi_n \vert \leq \eta_n, n =1,2,\dots,N,\label{eq:const2}\\[-3pt]
&~\setC_{\sf BR}:~ \sum_{k=1}^{K} \Vert \bold{\Phi} \bG\boldsymbol{w}_k \Vert_2^2 + \Vert \bm\phi  \Vert_2^2 \sigma_v^2 \leq P_A.\label{eq:const3}
	\end{align}
\end{subequations}
In problem~\eqref{eq:prob_form}, the constraint $\setC_{\sf BS}$ represents the total transmission power budget at the BS is no larger than $P_B$; one can instead consider the per-antenna power constraint
\begin{equation}\label{eq:per_ante}
    \| \bar{\bw}_m \|_2^2 \leq P_B/M,\quad m =1,\ldots, M,
\end{equation}
where $\bar{\bw}_m$ is the $m$th row of $(\boldsymbol{w}_1,\boldsymbol{w}_2,\dots,\boldsymbol{w}_K)$;
the constraint $\setC_{\sf RIS}$ restricts the reflecting gain at each RIS element no larger than a corresponding threshold $\eta_n$;
the constraint $\setC_{\sf BR}$ is the reflected power budget at the active RIS side.

The sum-rate maximization (SRM) problem in \eqref{eq:prob_form} is a basic problem in the active RIS system.
It resembles many system designs including secrecy rate maximization \cite{dong2022secure}, energy efficiency maximization \cite{niu2023energy}, and sub-connected array architecture \cite{liu2022subconnect,zhu2022subconnect}.
However, solving problem \eqref{eq:prob_form} is  difficult.
Problem~\eqref{eq:prob_form} is a non-convex program as the objective function and the constraint \eqref{eq:const3} are non-convex;
plus, the variables $\bw$ and $\bm \phi$ are coupled in both the objective function and constraint \eqref{eq:const3}.
In the existing literature, the most popular way is to apply block coordinate descent (BCD) and general purpose optimizers, such as CVX \cite{grant2014cvx}; see, e.g., \cite{zhang2023active,dong2022secure,liu2022subconnect,zhu2022subconnect,niu2023energy}.
This can incur prohibitively high computational complexity
when $M$ and $N$ are large, which happens in massive MIMO and large-scale RIS systems.
Our goal in this paper is to custom-build an efficient algorithm to tackle problem \eqref{eq:prob_form}.
As will be shown, our algorithm has a (semi) closed-form update in each step, and its efficiency will be demonstrated by simulation.

\section{Proposed Algorithm}

Our method has two key ingredients: 1) reformulating the objective function, and 2) dealing with the coupled constraint.

\subsection{Reformulation of the Objective Function}

We apply a reformulation to the objective function,
which is shown in the following fact.
\begin{Fact}\label{fact:WMMSE}
  Problem \eqref{eq:prob_form} is equivalent to the following problem
\begin{equation}\label{eq:WMMSE}
\begin{split}
  \min_{\bw, \bm \phi, \bm \bu, \bm \rho}  g(\bw, \bm \phi, \bu, \bm \rho)\quad
  {\rm s.t.}~\setC_{\sf BS},~\setC_{\sf RIS},~ \setC_{\sf BR},
  \end{split}
\end{equation}
where $\bm\rho\in \Rbb_{++}^K$, $\bu\in \Cbb^{K}$, and
\begin{equation*}
  \begin{split}
 g(\bw, \bm \phi, \bu, \bm \rho)=&~ \textstyle \sum_{k=1}^{K}  \rho_k F_k(\bw,\bm \phi,u_k) - \log (\rho_k),\\ 
  F_k(\bw,\bm \phi,u_k)  = &~ \textstyle \vert u_k \vert^2 \big( \sum_{i=1}^{K} \vert \bh_k^H \boldsymbol{w}_i \vert^2 + \Vert \bm f_k^H \bold{\Phi} \Vert_2^2 \sigma_v^2 \\ 
  & + \sigma_k^2 \big)  - 2{\rm{Re}}\big( u_k^* \bh_k^H \bw_k  \big) + 1. \\[-3pt]
  \end{split}
\end{equation*}
\end{Fact}

Fact~\ref{fact:WMMSE} can be shown in the same vein as the WMMSE method \cite{shi2011wmmse}. 
We shall omit the proof details.

%

\subsection{Precoding Algorithm}

In Fact 1, with the aid of the auxiliary variables $(\bu, \bm \rho)$, the reformulated problem \eqref{eq:WMMSE} can be efficiently tackled by the block successive upper bound minimization (BSUM) method~\cite{razaviyayn2013unified}.
To describe, let $(\bw^{\ell}, \bm \phi^{\ell}, \bm \bu^{\ell}, \bm \rho^{\ell})$ denote the solution at current iteration $\ell$.
The BSUM method takes the following updates
\begin{equation}\label{eq:BCD}
  \begin{split}
   \bu^{\ell+1} = &~\arg\min_{\bu} g(\bw^{\ell}, \bm \phi^{\ell}, \bm \bu, \bm \rho^{\ell});\\
   \bm \rho^{\ell+1} = &~\arg\min_{\bm \rho}g(\bw^{\ell}, \bm \phi^{\ell}, \bm \bu^{\ell+1}, \bm \rho);\\
   \bw^{\ell+1} = &~\arg\min_{\bw} g(\bw, \bm \phi^{\ell}, \bm \bu^{\ell+1}, \bm \rho^{\ell+1});\\
   \bm \phi^{\ell+1} = &~\arg\min_{\bm \phi} g(\bw^{\ell+1}, \bm \phi, \bm \bu^{\ell+1}, \bm \rho^{\ell+1}).\\[-3pt]
  \end{split}
\end{equation}

The solutions to the $\bu$ and $\bm \rho$ subproblems can be computed in closed forms, which are given by
\begin{equation*}
  \begin{split}
  u_k^{\ell+1} = &~ \frac{ \bh_k^H \bw_k^{\ell} }{ \sum_{i=1}^{K} \vert \bh_k^H \bw_i^{\ell} \vert^2 + \Vert \bm f_k^H \bm \Phi^{\ell} \Vert_2^2 \sigma_v^2 + \sigma_k^2}, ~k=1,\ldots, K,\\
  \rho_k^{\ell+1} =&~ (1-(u_k^{\ell+1})^* \bh_k^H \bw_k^{\ell} )^{-1}, ~k=1,\ldots, K.
  \end{split}
\end{equation*}

The $\bw$ and $\bm \phi$ subproblems do not have closed-form solutions.
We first consider the $\bm \phi$ subproblem, which takes the form
\begin{equation}\label{eq:phi_sub}
  \begin{split}
   \min_{\bm \phi} &~ {\bm\phi}^H \bQ \bm\phi - \Re\{ 2\bm\phi^H \bz\} \\[-3pt]
	\mbox{s.t.}	&~ \mathcal{C}_{\sf RIS}: ~\vert \phi_n \vert \leq \eta_n, ~ n =1,2,\dots,N,\\[-3pt]
	&~	\mathcal{C}_{\sf BR}:
		\bm\phi^H \bm\Lambda \bm\phi \leq P_{A},
  \end{split}
\end{equation}
where $\bW^{\ell+1}=\sum_{i=1}^{K}\bw_i^{\ell+1}(\bw_i^{\ell+1})^H$ and
\begin{equation*}
  \begin{split}
   \bm\Lambda  = &\textstyle  \sum_{k=1}^{K} {\rm{Diag}}(\bG\bw_k^{\ell+1}) {\rm{Diag}}(\bG\bw_k^{\ell+1})^H + \sigma_v^2 \bI_N,\\ 
    \bQ =& \textstyle \sum_{k=1}^{K} | u_k^{\ell+1} |^2 \rho_k^{\ell+1} {\rm{Diag}}(\bm f_k^H) \bG\bW^{\ell+1} \bG^H {\rm{Diag}}(\bm f_k) \\ 
    &~ \textstyle + \sigma_v^2\sum_{k=1}^{K} | u_k^{\ell+1} |^2 \rho_k^{\ell+1} {\rm{Diag}}(\bm f_k^H) {\rm{Diag}}(\bm f_k),\\ 
    \bz=& \textstyle \! \sum_{k=1}^{K} {\rm{Diag}}(\bm f_k^H) \bG \big( \rho_k^{\ell+1} (u_k^{\ell+1})^* \bw_k - \rho_k^{\ell+1} | u_k^{\ell+1} |^2 \bW^{\ell+1} \bh_k \big).\\[-5pt]
  \end{split}
\end{equation*}

We tackle the problem \eqref{eq:phi_sub} by the proximal distance algorithm \cite{keys2019proximal}, which recast the problem as
\begin{equation}\label{eq:phi_PDA}
  \begin{split}
   \min_{\bm \phi} &~ {\bm\phi}^H \bQ \bm\phi - \Re\{ 2\bm\phi^H \bz\} + \mu \cdot \mbox{dist}^2(\bm \phi,\setC_{\sf BR}\cap \setC_{\sf RIS} ),
  \end{split}
\end{equation}
where $\mbox{dist}(\bx, \setX) = \min_{\by\in \setX} \| \bx - \by \|_2$
is the distance function from a point $\bx$ to a set $\setX$, and $\mu>0$ is a given penalty parameter.
Problem \eqref{eq:phi_PDA} approaches problem \eqref{eq:phi_sub} when $\mu \rightarrow +\infty$.

Note that problem \eqref{eq:phi_PDA} takes a bilevel optimization form as the distance function is embedded within the main problem.
We majorize the distance function by
\[
\mbox{dist}(\bm \phi,\setC_{\sf BR}\cap \setC_{\sf RIS} ) \leq \| \bm \phi - \Pi_{\setC_{\sf BR}\cap \setC_{\sf RIS}}(\bm \phi^{\ell}) \|_2,
\]
where $\Pi_{\setX}(\bx) =\arg\min_{\by\in \setX} \|\bx -\by \|_2  $ is the projection of $\bx$ onto the set $\setX$.
Then, problem \eqref{eq:phi_PDA} can be majorized by
\begin{equation}\label{eq:phi_PDA2}
  \begin{split}
   \min_{\bm \phi} &~ {\bm\phi}^H \bQ \bm\phi - \Re\{ 2\bm\phi^H \bz\} + \mu  \| \bm \phi - \Pi_{\setC_{\sf BR}\cap \setC_{\sf RIS}}(\bm \phi^{\ell}) \|_2^2,
  \end{split}
\end{equation}
which has a closed form solution that is given by
\[
    \bm \phi^{\ell+1} = (\bQ + \mu \bI_N)^{-1}(\bz+ \mu \Pi_{\setC_{\sf BR}\cap \setC_{\sf RIS}}(\bm \phi^{\ell})).
\]

It remains to derive the projection $\Pi_{\setC_{\sf BR}\cap \setC_{\sf RIS}} (\bm \phi)$.
Interestingly, we found that the projection has a semi-closed form solution, as shown in Fact~\ref{fact:proj}.
The proof of Fact~\ref{fact:proj} is shown in the Appendix.
\begin{Fact}\label{fact:proj}
  For a fixed $\bw$, the projection $\hat{\bm \phi} = \Pi_{\setC_{\sf BR}\cap \setC_{\sf RIS}} (\bm \phi)$
  is given by
  \[
    \hat{\phi}_i = \begin{cases}
                \min ( |\phi_i| , \eta_i   ) \cdot e^{\jj\angle \phi_i} , & \mbox{if } \sum\limits_{i=1}^{N} \lambda_i \min ( a_i , \eta_i   )^2\leq P_{A}  \\
                 \min\big\{\frac{|\phi_i|}{1+\gamma \lambda_i  } ,\eta_i\big\} e^{\jj\angle \phi_i} , & \mbox{otherwise}.
                \end{cases}
  \]
for $i=1,\ldots, N$, where $\gamma \geq 0$ is chosen such that $\hat{\bm\phi}^H \bm\Lambda \hat{\bm\phi} = P_{A}$; $\lambda_i$ is the $i$th diagonal element of $\bm \Lambda$.
\end{Fact}

We tackle the $\bw$ subproblem by the same rationale.
The $\bw$ subproblem amounts to solve
\begin{equation}\label{eq:W_subprob}
  \begin{split}
    \min_{\bw} &~\textstyle
		\sum_{k=1}^{K}\bw_k^H \bA \bw_k -  \sum_{k=1}^{K}\Re \{ 2\bm b_k^H \bw_k \} \\ 
\mbox{s.t.} &~
		\textstyle\setC_{\sf BS}:   \sum_{k=1}^{K}\|\bw_k \|_2^2 \leq P_B,~\setC_{\sf BR}:\sum_{k=1}^{K}\bw_k^H \bm \Psi \bw_k \leq P,\\ 
  \end{split}
\end{equation}
where $\bA = \sum_{k=1}^{K} \rho_k^{\ell+1} | u_k^{\ell+1} |^2 \bh_k \bh_k^H$, $\bm b_k = \rho_k^{\ell+1} (u_k^{\ell+1})^* \bh_k$, $P = P_{A} -\|\bm\phi^{\ell}\|_{2}^2 \sigma_v^2 $, 
 and $ \bm\Psi = \bG^H (\bm\Phi^{\ell})^H \bm\Phi^{\ell} \bG$.
The set $\setC_{\sf BS}\cap \setC_{\sf BR}$ is an intersection of a ball and an ellipsoid, whose associated projection does {\it not} have an explicit solution.
To bypass this issue, we apply distance majorization independently for the two sets $\setC_{\sf BS}$ and $\setC_{\sf BR}$.
To be specific, we approximate problem \eqref{eq:W_subprob} by
\begin{equation}\label{eq:PDA1}
  \begin{split}
    \min_{\bw} &~\textstyle \sum_{k=1}^{K}\bw_k^H \bA \bw_k -  \sum_{k=1}^{K}\Re \{ 2\bm b_k^H \bw_k \} \\ 
&~+ \mu \cdot \mbox{dist}^2(\bw, \setC_{\sf BS}) + \mu \cdot\mbox{dist}^2(\bw, \setC_{\sf BR}).
  \end{split}
\end{equation}
Since $\setC_{\sf BS}$ and $\setC_{\sf BR}$ are convex and bounded, solving problem \eqref{eq:PDA1} will yield a solution $\bw$ that is close to the set $\setC_{\sf BS} \cap \setC_{\sf BR} $; see \cite[Proposition 2]{Lange2015PDA} for details.

Then, we majorize the distance function by
\begin{equation}\label{eq:dis_maj_w}
    \mbox{dist}(\bw, \setC_{\sf BS}) \leq \|\bw - \tilde{\bw}_{\sf BS}\|_2,~
    \mbox{dist}(\bw, \setC_{\sf BR}) \leq \|\bw - \tilde{\bw}_{\sf BR}\|_2,
  \end{equation}
where $\tilde{\bw}_{\sf BS}=\Pi_{\setC_{\sf BS}}(\bw)$, $\tilde{\bw}_{\sf BR}=\Pi_{\setC_{\sf BR}}(\bw)$  and
\begin{equation*}
  \begin{split}
     [\tilde{\bw}_{\sf BS}]_k = &\begin{cases}
                                  \bw_k, & \mbox{if } \| \bw \|_2^2 \leq P_B \\
                                 \sqrt{P_B}\frac{\bw_k}{\| \bw \|_2} , & \mbox{otherwise},
                                \end{cases}\\ 
    [\tilde{\bw}_{\sf BR}]_k  = &\begin{cases}
                                  \bw_k, & \mbox{if } \sum_{k=1}^{K}\bw_k^H \bm \Psi \bw_k \leq P \\
                                  (\bI_{M} + 2\nu \bm \Psi)^{-1} \bw_k , & \mbox{otherwise},
                                \end{cases}
  \end{split}
\end{equation*}
for $k=1,\ldots, K$, where $\nu>0$ ensures $ \sum_{k=1}^{K} [\tilde{\bw}_{\sf BR}]_k^H  \bm \Psi  [\tilde{\bw}_{\sf BR}]_k  = P$ and can be found by the one-dimensional bisection search.

We apply the distance majorization \eqref{eq:dis_maj_w} to problem \eqref{eq:PDA1}.
Solving the resulting problem leads to a closed-form solution
\[
    \bw_k^{\ell+1} =    (2\mu \bI_{M} +  \bA)^{-1} (\bb_k+ \mu[\tilde{\bw}_{\sf BS}]_k^{\ell}+ \mu [\tilde{\bw}_{\sf BR}]_k^{\ell} ),~k =1,\ldots, K.
\]

The whole algorithm is summarized in Algorithm~\ref{Alg:BCD}.
It is seen that each step of Algorithm~\ref{Alg:BCD} has a closed form, which contributes to the efficient iteration updates.
In addition, for the penalty parameter $\mu$, we initialize it by a small value and keep gradually increasing its value as the algorithm proceeds.
This strategy is known as {\it homotopy} optimization, which can help avoid bad local minima and achieve good performance; see, e.g., \cite{watson1989homotopy,eunlavy2005homotopy,shao2021binary}, for detailed discussions.
\begin{algorithm}[ht!]
\caption{An Efficient BSUM Algorithm for Problem \eqref{eq:WMMSE}}\label{Alg:BCD}
\begin{algorithmic}[1]
\STATE {\bf Input:}  initialization $\bw$, $\bm \phi$ and $\mu$; $\gamma>1$;
$\ell=1$.
\REPEAT

\STATE   $u_k^{\ell+1} =   \frac{ \bh_k^H \bw_k^{\ell} }{ \sum_{i=1}^{K} \vert \bh_k^H \bw_i^{\ell} \vert^2 + \Vert \bm f_k^H \bm \Phi^{\ell} \Vert^2 \sigma_v^2 + \sigma_k^2}, ~k=1,\ldots, K$;

  \STATE  $\rho_k^{\ell+1} =  (1-(u_k^{\ell+1})^* \bh_k^H \bw_k^{\ell} )^{-1}, ~k=1,\ldots, K$;

  \STATE     $ \bw_k^{\ell+1} =    (2\mu \bI_{M} +  \bA)^{-1} (\bb_k+ \mu[\tilde{\bw}_{\sf BS}]_k^{\ell}+ \mu [\tilde{\bw}_{\sf BR}]_k^{\ell} ),~k =1,\ldots, K$;

  \STATE $ \bm \phi^{\ell+1} = (\bQ + \mu \bI_{N})^{-1}(\bz+ \mu \Pi_{\setC_{\sf BR}\cap \setC_{\sf RIS}}(\bm \phi^{\ell}))$;

\STATE $\mu = \mu \cdot \gamma $;

\STATE $\ell= \ell+1$;
\UNTIL {some stopping criterion is satisfied.}
\end{algorithmic}
\end{algorithm}

We should discuss the complexity of Algorithm \ref{Alg:BCD}. 
It can be verified that the per-iteration complexity of Algorithm~\ref{Alg:BCD} is $\cO (M^3+N^3 + M^2N+ M^2K + N^2M+N^2K+MNK)$.
By comparison, the per-iteration complexity of the algorithm in \cite{zhang2023active,liu2022subconnect} is $\cO ( M^{4.5} + N^{4.5} + MNK )$, and the per-iteration complexity of the algorithm  in \cite{zhu2022subconnect} is $\cO (N^{4.5} + M^{4.5}K^{4.5} + M^2K+NK^2 )$.

\section{Simulation Results and Conclusion}

In this section, we provide simulation results to show the performance of our algorithm.
The simulation settings are as follows.
We set $r=8m$ and ${\rm{L}}=100m$ in Fig. \ref{fig:system_model}.
For the small-scale fading, we employ the Rician fading model for all channels, see (37) in \cite{zhang2023active}.
The large-scale fading of BS-user channel is set as ${\rm{PL}}=41.2+28.7{\rm{log}}(d)$, and all large-scale fading of BS-RIS channel and RIS-user channel are set as ${\rm{PL}}=37.3+22.0{\rm{log}}(d)$, where $d$ is the distance between two devices \cite{zhang2023active}.
We set $\sigma_k^2 = \sigma_v^2 = -80 {\rm{dBm}}$, $\eta_n = 8, \forall n$, and $K=16$.
The benchmark schemes are: 1) the algorithm in \cite{zhang2023active,liu2022subconnect}, implemented by its open source code\cite{sourcecode}, which applies BCD and CVX solver; 
and 2) the algorithm in \cite{zhu2022subconnect}, which applies BCD, alternating direction method of multipliers (ADMM), and CVX solver.
We stop all the algorithms under the same stopping criterion $\left| R_k(\bw^{\ell+1},\bm \phi^{\ell+1} ) -  R_k(\bw^{\ell},\bm \phi^{\ell} ) \right| \leq 10^{-4}$.
In addition to the total power constraint \eqref{eq:const1}, we also consider the design problem under the per-antenna power constraint \eqref{eq:per_ante}.
All the results are averaged over 200 independent simulation trials.
\begin{figure}[t]	
	\centering
	\includegraphics[width=7cm]{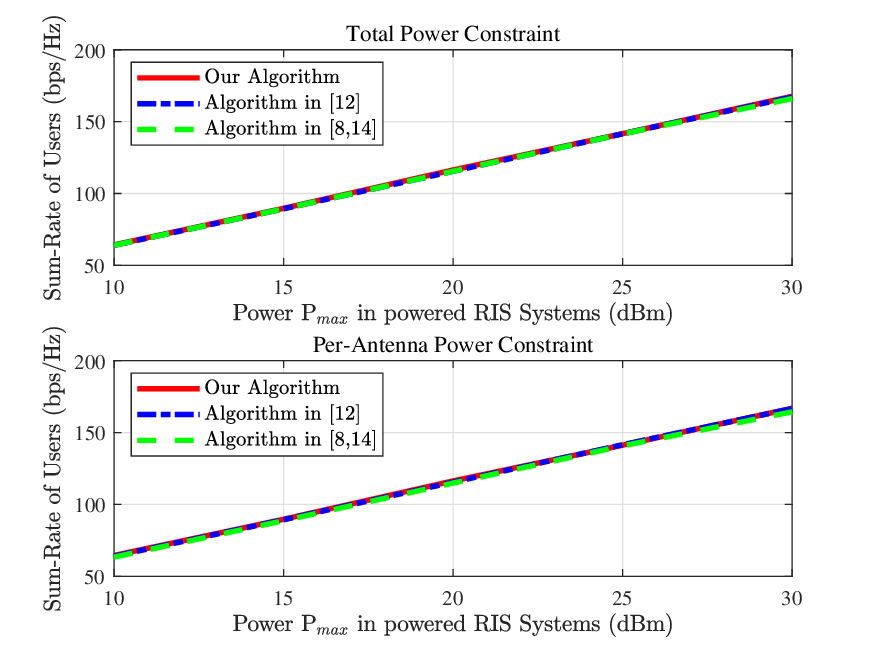}
	\caption{Multiuser sum rate of different algorithms.}
	\label{fig:sum-rate}
\end{figure}

\begin{figure}[t]	
	\centering
	{\includegraphics[width=7 cm]{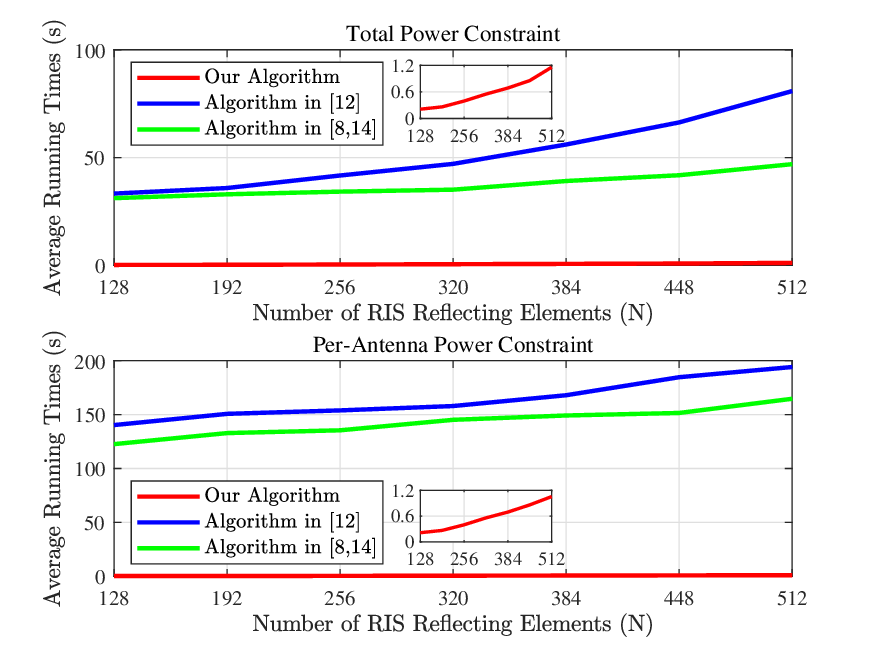}}
	\caption{Average runtime under different $N$; $M=256$.}
	\label{fig:rumtime_N}
\end{figure}

\begin{figure}[t]	
	\centering
	{\includegraphics[width=7 cm]{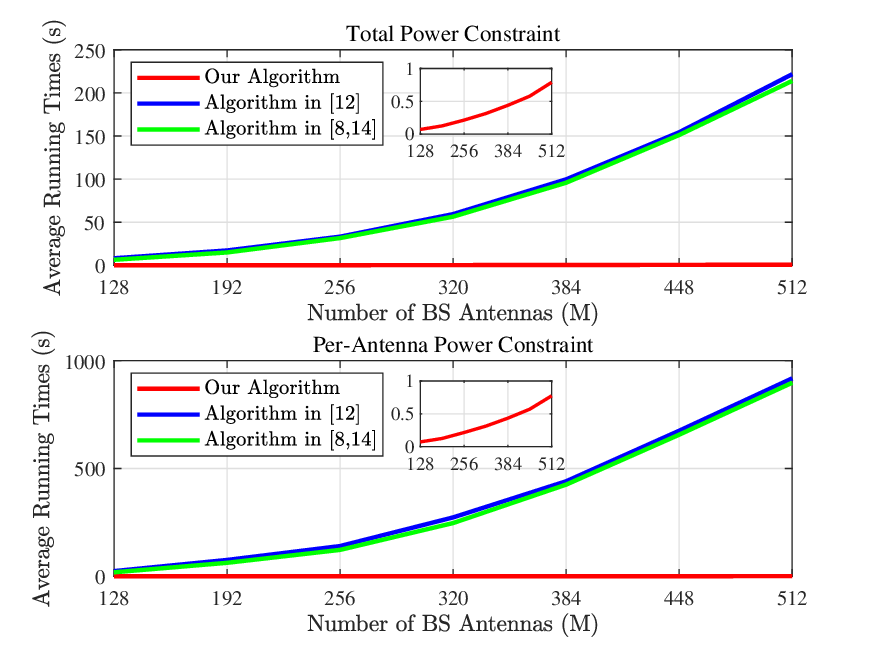}}
	\caption{Average runtime under different $M$; $N=128$.}
	\label{fig:runtime_M}
\end{figure}

In Fig.~\ref{fig:sum-rate}, we show the sum rate of all users achieved by different algorithms under different power budgets $P_{\max}$, where $P_{\max}$ denotes the overall power budget of the systems.
It is seen that all tested algorithms achieve comparable performance under both total power constraint and per-antenna power constraint.
However, their runtime performance can be quite different.
Figs.~\ref{fig:rumtime_N}-\ref{fig:runtime_M} show the average runtime of the considered algorithms under different MIMO-RIS system sizes.
In particular, Fig.~\ref{fig:rumtime_N} shows the average runtime over different number of RIS elements $N$ with fixed number of BS antenna elements $M$, while Fig.~\ref{fig:runtime_M} shows the average runtime over different number of antenna elements $M$ at the BS with fixed number of RIS elements $N$.
The overall power budget is set as $P_{\max} =30$ dBm; $P_A = 0.01 P_{max}$ and $ P_B = 0.99 P_{max} $.
First, it is seen that under all the tested settings our proposed algorithm is much faster than the other algorithms.
Second, it is seen that our proposed algorithm shows better scalability with respect to the problem size.
For instance, in Fig.~\ref{fig:runtime_M}, when $M=512$, the average runtimes of benchmark algorithms exceed 900 seconds, while our proposed algorithm consumes only about 0.8 seconds; in other words, our algorithm is more than 1,000 times faster.

To conclude, we have proposed an efficient BSUM method for the SRM problem in the active RIS system. 
The algorithm has simple closed-form solution in each step. 
The simulation results have shown that the proposed algorithm is significantly faster than existing algorithms. 

%


\begin{appendices}
	\section{Proof of Fact~2}
	Consider the projection problem
	\begin{equation}
		\begin{split}
		\hat{\bm\phi} &~= \arg	\min_{\by} \| \by - \bm \phi   \|_2^2 \\[-3pt]
			\mbox{s.t.}&~ | y_n |^2 \leq \eta_n^2, ~ n =1,2,\dots,N,~~ \textstyle\sum_{i=1}^{N}\lambda_i |y_i|^2 \leq P_{A}.
		\end{split}
	\end{equation}

	Here, note that $\bm \Lambda$ is a diagonal matrix and  $\lambda_i$ denote its diagonal elements.
	The associated Lagrangian function is given by
	\begin{equation}\label{eq:Lag_func}
	\textstyle {\cal L}_{\gamma, \bm \beta} (\by)=\| \by -\bm \phi \|_2^2 + \gamma \sum_{i=1}^{N}\lambda_i |y_i|^2 + \sum_{i=1}^{N} \beta_i |y_i|^2,
	\end{equation}
where $\gamma\geq 0$ and $\{\beta_i\}\geq 0$ are the dual variables.
	The Karush–Kuhn–\\Tucker (KKT) conditions are given by
	\begin{equation*}
	  \begin{split}
	   & \frac{\partial {\cal L}_{\gamma, \bm \beta}(\by)}{\partial y_i^*}= y_i -\phi_i + \gamma \lambda_i y_i + \beta_{i} y_i  = 0,~ \forall i,\\
	   & \textstyle \sum_{i=1}^{N}\lambda_i |y_i|^2 \leq  P_{A},~\gamma\geq  0,~ |y_i|^2\leq \eta_i^2,~ \beta_{i} \geq 0, \\
	    & \textstyle \gamma ( \sum_{i=1}^{N}\lambda_i |y_i|^2 - P_{A})=0, ~   \beta_{i}(|y_i|^2 - \eta_i^2) =0.
	  \end{split}
	\end{equation*}
	From the first line of KKT condition, we get 
	\begin{equation}\label{eq:kkt_grad}
	\textstyle y_i = \frac{1}{1+\gamma \lambda_i + \beta_i} \phi_i,
	\end{equation}
	which implies that $y_i$ and $\phi_i$ have the same angle.
	Also, the amplitude $|y_i|$ decreases as $\gamma$ and/or $\beta_i$ increase.
	It remains to specify the amplitude $|y_i|$'s.
	By the complementary slackness in the last line, we know that if $|y_i|<\eta_i$, then $\beta_i =0$ must hold.
	This, together with \eqref{eq:kkt_grad}, leads to
	\begin{equation}\label{eq:CS_1}
	 \textstyle   y_i = \min\big\{\frac{1}{1+\gamma \lambda_i  } |\phi_i|,\eta_i\big\} e^{\jj\angle \phi_i}.
	\end{equation}
	Again, by the complementary slackness with respect to $\gamma$, we know if $\sum_{i=1}^{N}\lambda_i |y_i|^2 <  P_{A}$, then $\gamma =0$ must hold and from $y_i = \min \{ |\phi_i|,\eta_i \} e^{\jj\angle \phi_i} $. Otherwise, $\gamma$ is chosen such that $\sum_{i=1}^{N}\lambda_i |y_i|^2 = P_{A}$.
	The proof is complete.
\end{appendices}


\bibliographystyle{IEEEbib}

\bibliography{refs}

\end{document}